\documentclass[runningheads]{llncs}
\usepackage[T1]{fontenc}
\usepackage{tikz}
\usetikzlibrary{arrows, positioning, shapes.geometric}
\usepackage{cite}
\usepackage{multirow}
\usepackage{amsmath}
\usepackage{amssymb}
\usepackage{booktabs}
\usepackage{amsfonts}
\usepackage{algorithm}
\usepackage{algorithmic}
\usepackage{graphicx}
\usepackage{subfigure}
\usepackage{pifont}
\usepackage{caption}
\usepackage{textcomp}
\usepackage{xcolor}
\usepackage{url}
\usepackage{hyperref}
\begin{document}
\title{DocAnnot - Accelerating the Creation of Key Information Extraction Datasets with GenAI-Powered Auto-Annotation}


%
\titlerunning{DocAnnot}
%
\author{Siddartha Reddy\inst{1} \and
Harikrishnan P M\inst{1} \and
Goutham Vignesh\inst{1}\ \and
Varun V\inst{1} \and
Vishal Vaddina\inst{2}}
\authorrunning{Siddartha Reddy et al.}
\institute{Phi Labs, Quantiphi India \\
\and
Phi Labs, Quantiphi Canada \\
\url{https://quantiphi.com} }

\maketitle              
\begin{abstract}
Key Information Extraction (KIE) from documents is a crucial task in many applications, but the creation of training datasets for KIE models is traditionally a time-consuming and expensive manual process. In this paper we introduce a novel Framework DocAnnot that significantly accelerates and reduces the cost of KIE dataset generation.  Our framework leverages a Large Vision Language Model (LVLM) for label value extraction, Optical Character Recognition (OCR) for text and bounding box detection, and a  novel Spatially Informed Contextual Matching (SICM) algorithm to accurately associate extracted values with their corresponding labels. SICM enhances label-value association by incorporating spatial relationships and proximity analysis alongside textual matching.  We evaluate our framework on two benchmark KIE datasets, CORD and SROIE, demonstrating its ability to automatically generate annotations with reasonable accuracy (achieving F1-scores of 0.679 and 0.846). Furthermore, we investigate the effectiveness of using the auto-annotated data for fine-tuning downstream KIE models.  While models trained on human-annotated data achieve superior performance, our experiments show that models trained exclusively on auto-annotated data still attain respectable performance levels (e.g., LayoutLMv3 achieving an F1-score of 0.6765 on CORD). These results demonstrate that while our framework significantly reduces reliance on manual annotation, it does not yet fully eliminate the need for human intervention. However, by automating the annotation process to a point where human reviewers can efficiently verify and refine the outputs, our system enables near-perfect annotations with far greater efficiency than manual annotation from scratch. This approach offers substantial time and cost savings while optimizing the balance between automation and accuracy, making it particularly valuable for resource-constrained settings and rapid model prototyping.


\end{abstract}
\vspace{-30pt} 


\newcommand\blfootnote[1]{%
  \begingroup
  \renewcommand\thefootnote{}\footnote{#1}%
  \addtocounter{footnote}{-1}%
  \endgroup
}

\blfootnote{\small Published in ICDAR 2025. DOI: \href{https://doi.org/10.1007/978-3-032-04624-6_33}{10.1007/978-3-032-04624-6\_33}}

\section{Introduction}

Key Information Extraction (KIE) \cite{bensch2021keyinformationextractiondocuments} from documents is a fundamental task in Intelligent Document Processing, with applications spanning automated data entry, document understanding, robotic process automation, and knowledge base construction. KIE is the process of automatically identifying and extracting specific pieces of information typically predefined keys or entity classes from unstructured document sources, and transforming them into a structured format \cite{skalicky2022business}. The effectiveness of these systems is critically dependent on the availability of large, high-quality, annotated datasets \cite{naparstek2024kvp10k}. However, the traditional method of creating such datasets involves manual annotation by human experts, a process that is inherently time-consuming, labor-intensive, and expensive \cite{goel2023llms}. This manual annotation bottleneck significantly hinders the development, deployment, and scalability of KIE models, limiting their adaptability to new document types and increasing overall costs.

Existing approaches to address this challenge fall into several categories, each with its own limitations. Manual annotation, while capable of producing high-quality data, is inherently slow and costly. Template-based methods, which rely on pre-defined rules and structures, are inflexible and struggle to adapt to documents with varying layouts or new, unseen document types.

Active learning attempts to optimize human annotation by selecting the most informative samples for labeling \cite{rouzegar2024enhancingtextclassificationllmdriven}, but it still requires substantial human involvement. Recent advancements in Large Vision Language Models (LVLM) \cite{achiam2023gpt,team2023gemini,TheC3,liu2023visual,bai2025qwen2}  have demonstrated impressive capabilities in understanding and reasoning about visual information combined with text, but these models are not yet finely tuned for the specific requirements of structured information extraction from documents, and may hallucinate without proper grounding \cite{zhang2024documentparsingunveiledtechniques}. Thus, a significant gap persists: there is a critical need for a fully automated, precise, and flexible approach to generate KIE datasets, where accurately assigning the labels to entities is essential.

To address this need, our paper introduces a novel Multimodal GenAI Powered Automated Annotation Framework (DocAnnot) to completely automate the KIE dataset creation process. Our framework leverages the strengths of three key components: an LVLM for label value extraction, Optical Character Recognition (OCR) for precise text and bounding box detection, and a novel Spatially Informed Contextual Matching (SICM) algorithm that intelligently associates extracted values by LVLM with their corresponding labels. The SICM algorithm is a critical contribution that effectively bridges the semantic understanding of the LVLM with the spatial information provided by the OCR. This approach is inspired by recent successes in leveraging Large Language Models (LLMs) for data augmentation and knowledge distillation \cite{goel2023llms,kim-etal-2024-dockd}, but it is specifically tailored to the challenges of structured information extraction from visually rich documents.

We evaluate our proposed framework on two widely-used benchmark KIE datasets, CORD \cite{park2019cord} and SROIE \cite{Huang_2019}, demonstrating its ability to generate annotations with reasonable accuracy, approaching, though not fully matching, the quality of human annotations. Critically, we also demonstrate that KIE models (LayoutLMv3 \cite{10.1145/3503161.3548112} and UDOP \cite{tang2023unifyingvisiontextlayout}) fine-tuned exclusively on data auto-annotated by our framework achieve respectable performance levels. This signifies that functional KIE models can be developed without any direct human intervention, offering a substantial reduction in the time and cost associated with KIE dataset creation.

Our key contributions are:
\begin{enumerate}
    \item A fully automated framework for KIE dataset generation.
    \item A novel Spatial-Informed Contextual Matching (SICM) algorithm that integrates LVLM and OCR outputs for accurate label assignment.
    \item Demonstrated effectiveness of the framework in benchmark KIE datasets.
    \item Validation of the auto-annotated data for training KIE models.
\end{enumerate}

\section{Related Work}

\subsection{Key Information Extraction (KIE) Models and Datasets}
Early approaches of KIE often relied on rule-based systems \cite{article} or template matching, which lacked generalizability to diverse document layouts. The introduction of the Transformer architecture \cite{vaswani2023attentionneed} revolutionized the field, offering significant improvements over earlier deep learning approaches that used models like CNNs \cite{oshea2015introductionconvolutionalneuralnetworks} and LSTMs \cite{10.1162/neco.1997.9.8.1735}. Transformer-based models, such as BERT \cite{devlin2019bertpretrainingdeepbidirectional}, demonstrated superior performance on various NLP tasks. This led to the development of specialized models for document understanding, such as LayoutLM \cite{Xu_2020}, LayoutLMv2 \cite{xu2022layoutlmv2multimodalpretrainingvisuallyrich}, and LayoutLMv3 \cite{10.1145/3503161.3548112}, which incorporate both textual and layout information, and, in later versions, visual features. LayoutLM \cite{Xu_2020} and its successors pre-train on large document corpora, using masked language modeling and other objectives to learn rich representations that capture the interplay between text, layout, and image. UDOP (Unified Document Processing) \cite{tang2023unifyingvisiontextlayout} represents a further step - unifying text, image, and layout modalities in a single model, and handling a variety of understanding and generation tasks. These models, which we also employ in our experiments, form the state-of-the-art in KIE. To train and evaluate these KIE models, several benchmark datasets have been created, including FUNSD \cite{jaume2019funsddatasetformunderstanding}, CORD \cite{park2019cord}, SROIE \cite{Huang_2019}, and DocVQA \cite{mathew2021docvqadatasetvqadocument}. These datasets provide a common ground for comparing different approaches and measuring progress in the field.



\subsection{Generative AI for Document Annotation}

The advent of LLMs and LVLMs has opened up new possibilities for automated data annotation. Recent works have explored using LLMs for data augmentation and knowledge distillation in various NLP tasks. Some approaches have used LLMs to guide the annotation of unlabeled data or to distill reasoning capabilities into smaller KIE models \cite{goel2023llms,kim-etal-2024-dockd}. Among recent efforts, DocKD \cite{kim-etal-2024-dockd} is, to the best of our knowledge, the only prior work that attempts a similar task. DocKD formulates KIE as a visual question-answering problem, requiring a separate LLM query for each label-value pair, resulting in high computational overhead and use to fine-tune a non-public DocFormerv2 model \cite{appalaraju2024docformerv2}. In contrast, our approach generates complete KIE annotations for an entire document image in a single LLM call, significantly reducing latency and cost with the novel SCIM algorithm.



\begin{figure}[htbp]
    \centering
    \includegraphics[width=\textwidth]{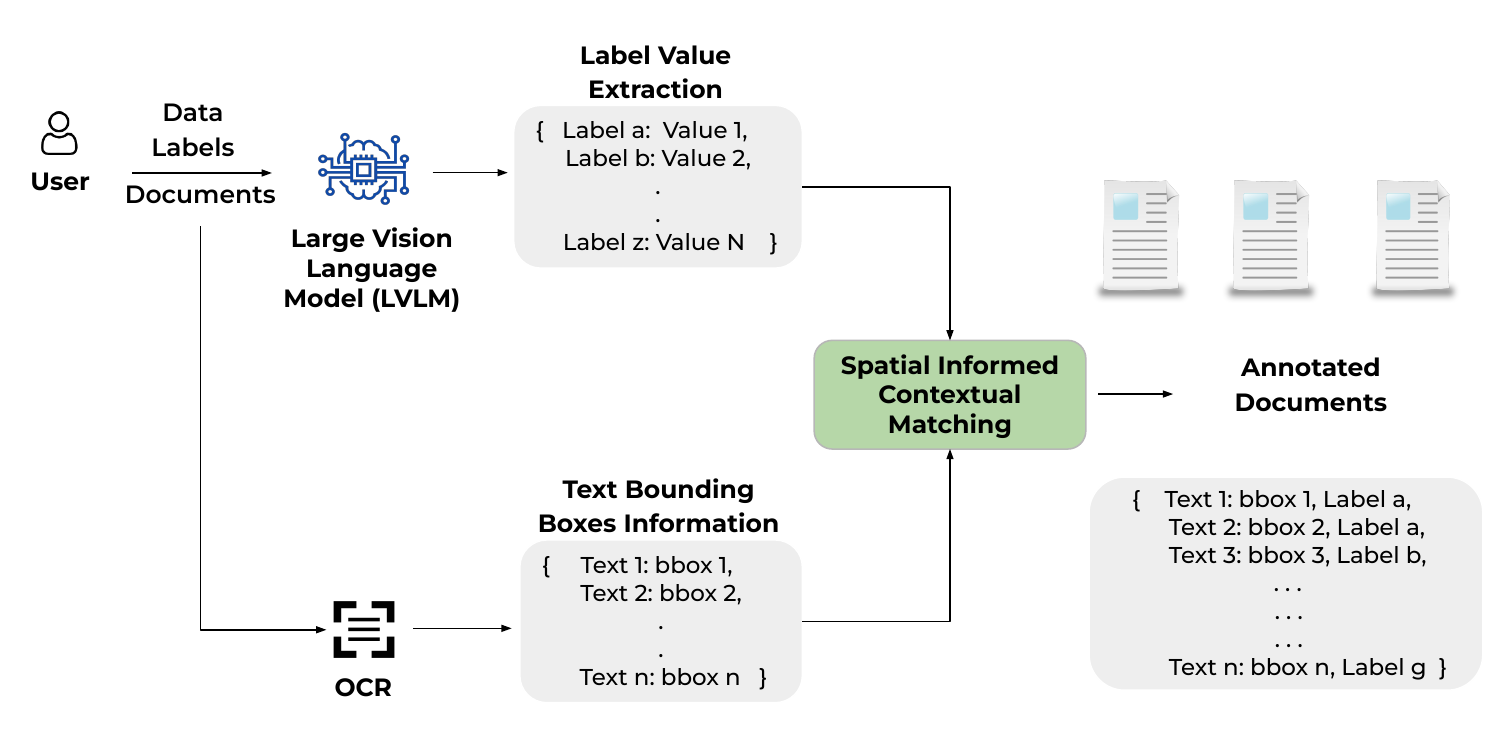}
 
    \caption{The DocAnnot framework for automated KIE dataset generation. Input documents and corresponding data labels are provided to the system. A Large Vision Language Model (LVLM) extracts label-value pairs, while Optical Character Recognition (OCR) extracts text and bounding box information. The Spatially Informed Contextual Matching (SICM) module then integrates these outputs to produce annotated documents, associating text spans with their corresponding labels.}
    \label{fig:architecture}
\end{figure}

\section{Methodology: Multimodal GenAI Powered Automated Annotation Framework}

\subsection{Overview of the Framework}


We introduce a novel framework for automated document annotation, designed to address the challenges of high cost and time associated with manual annotation in document KIE tasks.  As illustrated in Fig. \ref{fig:architecture}, the framework follows a multi-stage approach that integrates: 

\begin{itemize}

    \item \textbf{Large Vision Language Model (LVLM):} Utilized to extract label-value pairs from documents by understanding both textual and visual context.

    \item \textbf{Optical Character Recognition (OCR):} Employed to convert document images into machine-readable text.

    \item \textbf{Spatially Informed Contextual Matching (SICM) Algorithm:} A novel technique that enhances key information extraction by leveraging spatial and contextual cues.

\end{itemize}


\subsection{Automated KIE Dataset Generation: Merging LVLM Insights with OCR with SCIM}

For KIE models like LayoutLMv3, UDOP and similar, the training dataset is to be structured such that each document is provided as an image, and an accompanying OCR system extracts a sequence of text segments along with their corresponding bounding boxes. In addition, each text segment is assigned an entity label from a predefined set \(L = \{l_1, l_2, \ldots, l_m\}\). Once trained, the model is expected to predict a label for every word it receives as input from the OCR as a token classification problem \cite{Cao2023GenKIERG}. This structured format, which comprises the document image, OCR text with bounding boxes, and associated entity labels, forms the basis for our automated annotation framework.
\begin{figure}[htbp]
    \centering
    \includegraphics[width=\textwidth]{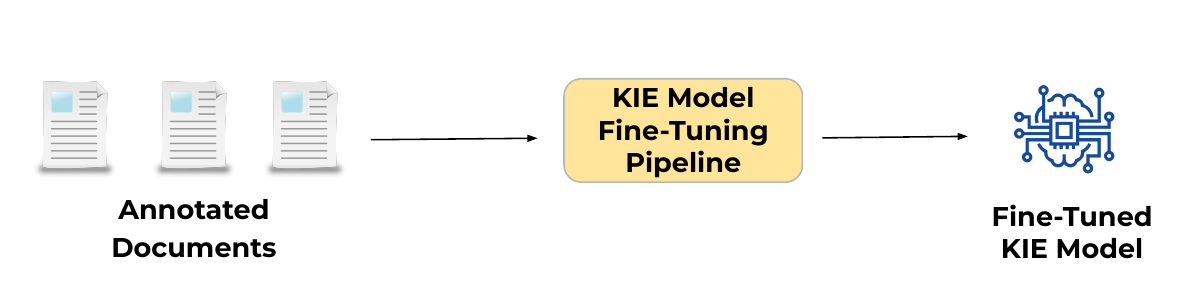}
 
    \caption{Fine-tuning a KIE model using the output of the DocAnnot framework. Annotated documents from DocAnnot serve as training data for the KIE model fine-tuning pipeline.}
    \label{fig:finetuning_pipeline}
\end{figure}

\subsubsection{LVLM-based Label Value Extraction:} The process begins with a collection of user-provided documents, \(D = \{d_1, d_2, \ldots, d_n\}\), and a set of user-defined entity labels L , relevant to the KIE task. An LVLM, denoted as \(\Phi_{LVLM}\), extracts label-value pairs from these documents. Importantly, for each label, the LVLM outputs not only the corresponding value but also the actual key text as it appears in the document. This can be represented as a mapping:
    \[
    \{\text{label}: \{\text{actual key text in document}: \text{value}\}\}
    \]
    For instance, a label-value pair may appear as \(\text{'sub\_total.subtotal\_price'}: \{\text{'SUBTOTAL'}: \text{'10,000'}\}\). In this context, the extracted key text (e.g., “SUBTOTAL”) is denoted as \(t_k\) and will be used in subsequent spatial matching.
The overall extraction is expressed as:
    \[
    LV = \Phi_{LVLM}(D, L)
    \]
    where \(LV\) is the set of label-value pairs containing both the value and the key text.

\subsubsection{Text and Bounding Box Extraction with OCR:} An OCR engine, denoted as \(\Phi_{OCR}\), processes the same set of documents \(D\). The OCR engine performs:
    \begin{enumerate}
        \item \textbf{Text detection with Bounding Box :} Determines the precise spatial coordinates (bounding boxes) for each text segment.
        \item \textbf{Text recognition:} Recognizes all detected text instances in the document.
    \end{enumerate}
The OCR output is represented as:
    \[
    TB = \Phi_{OCR}(D) = \{ (t_1: b_1), (t_2: b_2), \ldots, (t_n: b_n) \}
    \]
    where each \(b_i\) corresponds to the bounding box for text segment \(t_i\).
\subsubsection{Spatially Informed Contextual Matching (SICM):}

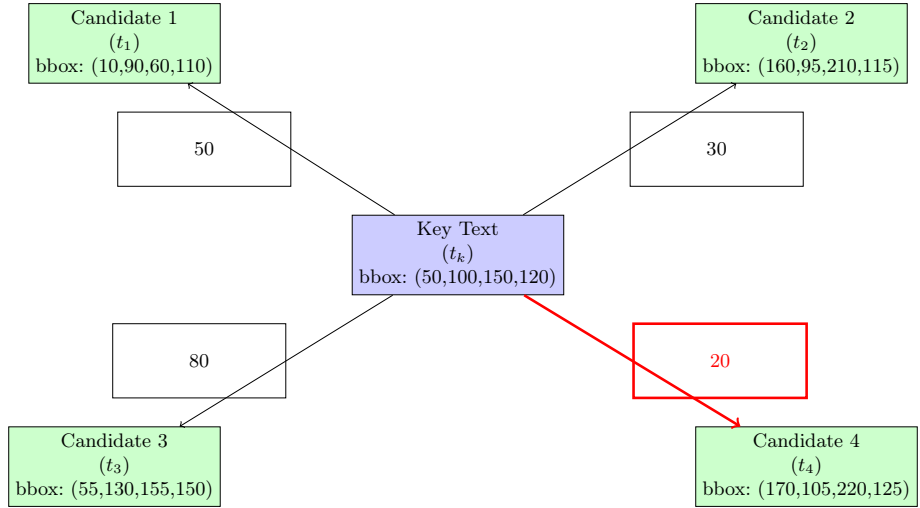
\begin{figure}[htbp]
    \centering
    \resizebox{\textwidth}{!}{
    \begin{tikzpicture}[
      node distance=3cm,
      every node/.style={draw, rectangle, minimum width=2.8cm, minimum height=1.2cm, align=center},
      edge from parent/.style={draw, -latex, thick}
    ]
      \node[fill=blue!20] (key) {Key Text\\($t_k$)\\\footnotesize{bbox: (50,100,150,120)}};

      \node[fill=green!20, above left=of key] (cand1) {Candidate 1\\($t_1$)\\\footnotesize{bbox: (10,90,60,110)}};
      \node[fill=green!20, above right=of key] (cand2) {Candidate 2\\($t_2$)\\\footnotesize{bbox: (160,95,210,115)}};
      \node[fill=green!20, below left=of key] (cand3) {Candidate 3\\($t_3$)\\\footnotesize{bbox: (55,130,155,150)}};
      \node[fill=green!20, below right=of key] (cand4) {Candidate 4\\($t_4$)\\\footnotesize{bbox: (170,105,220,125)}};

      \draw[->] (key) -- node[midway, left] {50} (cand1);
      \draw[->] (key) -- node[midway, right] {30} (cand2);
      \draw[->] (key) -- node[midway, left] {80} (cand3);
      \draw[->, very thick, red] (key) -- node[midway, right] {20} (cand4);
    \end{tikzpicture}
    }
    \caption{Disambiguation using bounding box proximity. The central element (\(t_k\)) represents the key text and its bounding box, while surrounding elements are candidate OCR segments. The red line indicates the shortest Euclidean distance, leading to the selection of the closest candidate.}

    \label{fig:graph}
\end{figure}

The SICM algorithm, denoted as \(\Phi_{\text{SICM}}\), is the core innovation of our framework. Its objective is to reconcile the outputs from the LVLM and OCR modules by assigning the correct labels from \(L\) to the OCR-extracted text segments. Recall that the LVLM produces label-value pairs \(LV\) which include the actual key text, and the OCR module outputs a set of text segments with their bounding boxes \(TB\). The SICM algorithm operates in two hierarchical steps:

\begin{enumerate}
    \item \textbf{Textual Similarity Filtering:}\\
    For each label-value pair \((l_i, v_i)\) in \(LV\) and every OCR text segment \(t_j\) (with bounding box \(b_j\)) in \(TB\), a textual similarity score is computed:
    \[
    Similarity(v_i, t_j)
    \]
    using a metric such as the Levenshtein distance. Candidates are then filtered based on this score (e.g., by selecting those exceeding a predefined similarity threshold or by choosing the top-N candidates). Denote this filtered set as \(TB'(v_i) \subset TB\).

    \item \textbf{Spatial Disambiguation via Nearest Neighbor:}\\
    In scenarios where the same value (e.g., “\$3.50”) appears multiple times, textual similarity alone may be insufficient to resolve ambiguity. To address this:
    \begin{itemize}
        \item The extracted key text \(t_k\) (e.g., “SUBTOTAL”), accompanied by its bounding box \(b_k\) and obtained during the LVLM stage, serves as a spatial reference.
        \item For each candidate OCR text segment in \(TB'(v_i)\), the Euclidean distance \(d(b_k, b_j)\) is computed between its bounding box \(b_j\) and the bounding box of the key text \(b_k\) provided by the LVLM. These distances are used to identify the candidate closest to the key text.
        \item The candidate with the smallest distance to \(b_k\) is then selected:
        \[
        b_{j^*} = \arg\min_{b_j \in TB'(v_i)} d(b_k, b_j)
        \]
    \end{itemize}
\end{enumerate}
To illustrate the disambiguation process using bounding box information, Fig.~\ref{fig:graph} provides a schematic overview. In this figure, the central element corresponds to the key text (\(t_k\)) extracted by the LVLM along with its bounding box, and the surrounding elements represent candidate OCR text segments with their respective bounding boxes. The lines between them are labeled with the Euclidean distances between the bounding box centers. The line highlighted in red indicates the candidate closest to the key text, which is selected for label-value assignment.
\\
\\
For each label-value pair \((l_i, v_i)\), the label assignment is defined as:
\[
Assignment(l_i, v_i) = (t_{j^*}, b_{j^*})
\]
meaning that the OCR text segment corresponding to \(b_{j^*}\) is selected for label \(l_i\).

Thus, the SICM function is mathematically formulated as:
\[
\Phi_{\text{SICM}}(LV, TB) = \{ Assignment(l_i, v_i) \;:\; l_i \in L \}
\]
and when applied over all label-value pairs, the final annotation is directly obtained as:
\[
A = \{ (t_{j^*}, b_{j^*}, l_i) \;:\; l_i \in L \}.
\]

This multistage process leverages both textual similarity and spatial reasoning to robustly resolve ambiguities in key-value linking, especially in cases where multiple instances of the same value occur. Such annotated data, combined with the corresponding document images, constitute the training data format expected by state-of-the-art KIE models such as LayoutLMv3, UDOP, and similar models.

\section{Experiments}

This section details the experiments conducted to evaluate the performance of the proposed Multimodal GenAI Powered Automated Annotation Framework.  We present two sets of experiments: (1) assessing the framework's ability to automatically annotate documents, and (2) evaluating the effectiveness of using the auto-annotated data for fine-tuning smaller models.

\subsection{Auto Annotation of the Documents}

This experiment aims to quantify the accuracy of the framework in automatically annotating documents compared to human-annotated ground truth data.

\textbf{Datasets:}  We utilized two publicly available datasets commonly used in KIE experiments:
\begin{enumerate}
    \item \textbf{CORD: }Consolidated Receipt Dataset \cite{park2019cord} for Post-OCR
Parsing is a key information extraction dataset with 30
labels under 4 categories such as "total" or "subtotal". It has
1,000 receipt samples. The train, validation, and test splits
contain 800, 100, and 100 samples respectively.

    \item \textbf{SROIE:} Scanned Receipts OCR \cite{Huang_2019} and Information Extraction dataset is a benchmark dataset introduced during the ICDAR 2019 competition. It comprises 973 scanned receipt images, divided into training and test sets, with 626 images for training and 347 images for testing. Each receipt is annotated with four key fields: company, date, address, and total.
    
\end{enumerate}

\subsubsection{Experimental Setup:}

\begin{enumerate}
    \item \textbf{Evaluation Set Size}: For each dataset, we used the standard test split for the evaluation.
    \item \textbf{LVLMs:} We employed two state-of-the-art LVLMs (Claude Sonnet 3.5 \cite{TheC3}, Gemini 1.5 Pro \cite{team2023gemini}) to extract label-value information within our framework.
    \item \textbf{Prompt Design:} We used both models in zero-shot setting with structured prompts, e.g., \textit{``Extract the following fields from the document: Vendor, Invoice Date, Total. Respond strictly using the following JSON schema:''} \texttt{\{"entity\_key": \{"entity\_key\_phrased\_in\_the\_document": "extracted value"\}\}}. This prompt format ensures compatibility with our evaluation pipeline.

    \item \textbf{OCR Configuration:} We utilized the OCR-extracted text and bounding box information provided with each dataset (CORD and SROIE). 

    \item \textbf{Evaluation Metrics:} Annotation performance was evaluated at the token level.  For each token in the OCR-extracted text sequence, we determined whether it was correctly assigned to the relevant label according to the ground-truth annotations.  We quantified the performance using standard classification metrics: Precision, Recall, and F1-score.
    

\end{enumerate}

\subsubsection{Results:}
\mbox{}\\[0.5em]
Table \ref{tab:auto_annotation_results_comparison} summarizes the auto-annotation performance of our proposed framework, DocAnnot, using Claude Sonnet 3.5 and Gemini 1.5 Pro as the underlying LVLMs, on the CORD and SROIE datasets.
\begin{table}[h!]
\centering
\caption{The table illustrates the performance of Claude Sonnet 3.5 and Gemini 1.5 Pro on the auto-annotation task, evaluated on the CORD and SROIE datasets. Performance is measured using Precision, Recall, and F1-score, calculated at the token level.}
\label{tab:auto_annotation_results_comparison}
\resizebox{\textwidth}{!}{  
\begin{tabular}{|l|c|c|c|c|c|c|}
\hline
\multirow{2}{*}{Model} & \multicolumn{3}{c|}{CORD} & \multicolumn{3}{c|}{SROIE} \\
\cline{2-7}
 & Precision & Recall & F1 Score & Precision & Recall & F1 Score \\
\hline
DocAnnot\textsubscript{Claude Sonnet 3.5}  & \textbf{0.707} & \textbf{0.704} & \textbf{0.679} & \textbf{0.886} & \textbf{0.853} & \textbf{0.846} \\
DocAnnot\textsubscript{Gemini 1.5 Pro} & 0.658 & 0.638 & 0.626 & 0.766 & 0.795 & 0.768 \\
\hline
\end{tabular}
}
\end{table}

On the CORD dataset when using Claude Sonnet 3.5 as the LVLM, DocAnnot achieved a Precision of 0.707, a Recall of 0.704, and an F1-score of 0.679.  With Gemini 1.5 Pro as the LVLM, DocAnnot achieved a lower Precision of 0.658, a Recall of 0.638, and an F1-score of 0.626.  These results demonstrate that DocAnnot performs more effectively with Claude Sonnet 3.5 on the CORD dataset. Similarly on the SROIE dataset DocAnnot, leveraging Claude Sonnet 3.5 as its LVLM, achieved a Precision of 0.886, a Recall of 0.853, and an F1-score of 0.846.  Using Gemini 1.5 Pro as the LVLM, DocAnnot achieved a Precision of 0.766, a Recall of 0.795, and an F1-score of 0.768. While performance was strong with both LVLMs, DocAnnot with Claude Sonnet 3.5 maintained superior performance.

Across both datasets, DocAnnot consistently achieved higher auto-annotation accuracy (as measured by the F1-score) when using Claude Sonnet 3.5 as its LVLM compared to Gemini 1.5 Pro. This superior performance with Claude Sonnet 3.5 indicates a better balance between correctly identifying relevant tokens (Precision) and capturing all relevant tokens (Recall). The observed performance difference between the CORD and SROIE datasets, with higher scores on SROIE for DocAnnot with both LVLMs, suggests potential differences in dataset characteristics or complexities that influence the framework's ability to perform auto-annotation. Further investigation into these dataset-specific factors is warranted.

\subsection{Fine-Tuning Smaller Models with Auto-Annotated Data}

This experiment evaluates the effectiveness of using data auto-annotated by DocAnnot for training downstream KIE models.  The core question we address is: Can data automatically annotated by DocAnnot replace or augment traditionally and expensively acquired human-annotated data for training KIE models?

\subsubsection{Experimental Setup: }

\begin{enumerate}

\item \textbf{Dataset:} The CORD dataset \cite{park2019cord} was used for all experiments. 
\item \textbf{DocAnnot Configuration:} Claude Sonnet 3.5 was employed within the DocAnnot framework to generate auto-annotated data used for training downstream KIE models.
\item \textbf{Downstream KIE Models:} LayoutLMv3, and UDOP are considered for KIE task finetuning. We trained separate instances of both LayoutLMv3 and UDOP with human annotated data and annotations generated by DocAnnot. All trained models were then evaluated on the standard CORD test set to ensure a consistent comparison.
\end{enumerate}

\subsubsection{Results: }
\mbox{}\\[0.5em]
Table \ref{tab:finetuning_results} presents the fine-tuning performance of LayoutLMv3 \cite{10.1145/3503161.3548112} and UDOP \cite{tang2023unifyingvisiontextlayout} on the CORD data set, comparing the training results with human-annotated data versus auto-annotated data.  We also include the performance of DocFormerv2large \cite{appalaraju2023docformerv2localfeaturesdocument} using DocKD-Annotated data \cite{kim-etal-2024-dockd} as a baseline.

\begin{table}[h!]
\centering
\caption{Fine-Tuning Performance on the CORD Dataset: Comparison of LayoutLMv3 and UDOP trained on human-annotated vs. auto-annotated data, and a DocFormerv2\textsubscript{large} \cite{appalaraju2023docformerv2localfeaturesdocument} baseline trained on Doc-KD \cite{kim-etal-2024-dockd} annotated data.}
\label{tab:finetuning_results}
\resizebox{\textwidth}{!}{  
\begin{tabular}{|l|l|c|c|c|}
\hline
Model & Training Data & Precision & Recall & F1 Score \\
\hline
DocFormerv2\textsubscript{large} & DocKD-Annotated   & -      & -      & 0.615 \\
LayoutLMv3                       & Human-Annotated   & 0.9086 & 0.9198 & 0.9141 \\
LayoutLMv3                       & Auto-Annotated    & 0.7217 & 0.6367 & 0.6765 \\
UDOP                             & Human-Annotated   & 0.912  & 0.928  & 0.921  \\
UDOP                             & Auto-Annotated    & 0.7364 & 0.6424 & 0.6861 \\
\hline
\end{tabular}
}
\end{table}




Despite the performance gap with human annotated data, a key finding is that fine-tuning exclusively on auto-annotated data still yields a respectable level of performance. Notably, LayoutLMv3 achieves an F1-score of 0.6765 and UDOP achieves 0.6861 both surpassing the DocFormerV2large baseline (F1-score of 61.5), which also uses auto-annotated data (from DocKD). This highlights the quality of data produced by DocAnnot, enabling the training of effective KIE models without human annotation. Although the UDOP score marginally exceeds the original auto-annotation F1 of 0.679, multiple fine-tuning runs (ranging from 0.669 to 0.681) confirm that this behavior is consistent with student–teacher learning setups. Fine-tuning helps smooth over noisy supervision \cite{muller2019does}, allowing the model to generalize slightly better than the annotation source. Moreover, outperforming another auto-annotation method further underscores the value of our framework, offering a substantial reduction in annotation effort and expense while maintaining competitive performance. This marks a significant step toward more efficient and scalable KIE model development.

\section{Ablation Study} 
\subsection{Isolating the Effect of SICM}
To quantify the contribution of our proposed SICM algorithm, we conducted an ablation study comparing the performance of our automated annotation framework with and without SICM. This analysis focuses on the automated annotation task, which evaluates F1, precision and recall in the SROIE and CORD datasets. We utilized Claude 3.5 Sonnet as the LVLM for label-value extraction in both configurations. The “with SICM” setting corresponds exactly to the full system reported in Table~\ref{tab:auto_annotation_results_comparison}, including both textual similarity filtering and spatial disambiguation. In contrast, the “without SICM” baseline disables spatial filtering and relies solely on fuzzy string matching for alignment. Both variants use the same LVLM outputs, ensuring that the effect of SICM is isolated in the comparison.
\begin{table}[htbp]
\centering
\caption{Evaluating the impact of Spatially Informed Contextual Matching (SICM) on Auto-Annotation. The table shows improved Precision, Recall, and F1 Score on CORD and SROIE using Claude 3.5 Sonnet as the LVLM.}
\label{tab:ablation}
\resizebox{\textwidth}{!}{  
\begin{tabular}{|l|c|c|c|c|c|c|}
\hline
\multirow{2}{*}{Method} & \multicolumn{3}{c|}{SROIE} & \multicolumn{3}{c|}{CORD} \\
\cline{2-7}
 & Precision & Recall & F1 Score & Precision & Recall & F1 Score \\
\hline
DocAnnot\textsubscript{without SICM} & 0.846 & 0.805 & 0.814 & 0.580 & 0.560 & 0.571 \\
DocAnnot\textsubscript{with SICM}    & \textbf{0.886} & \textbf{0.853} & \textbf{0.846} & \textbf{0.707} & \textbf{0.704} & \textbf{0.679} \\
\hline
\end{tabular}}
\end{table}

As evidenced by the results in Table \ref{tab:ablation}, the inclusion of SICM significantly enhances the performance of the DocAnnot framework. For the SROIE dataset, precision increased from 0.846 to 0.886, while recall improved from 0.805 to 0.853. The improvements are even more pronounced in the CORD dataset, where precision rose from 0.580 to 0.707 and recall from 0.560 to 0.704. F1 scores also followed a similar upward trend in both datasets. These results clearly demonstrate the effectiveness of SICM in enhancing the model’s ability to accurately associate labels with their corresponding values, leading to more reliable annotations.

\subsection{Hybrid Annotation with Limited Human Labels}

To assess the utility of DocAnnot in practical human-in-the-loop annotation pipelines, we conducted a hybrid annotation experiment on the CORD dataset.  
UDOP was fine-tuned using DocAnnot autoannotated data, combined with a small fraction of human-labeled ground truth samples (\(10\%\), \(20\%\), and \(30\%\) of the training split).  
As shown in Table~\ref{tab:hybrid_annotation}, F1 scores improved from \(0.6861\) (auto only) to \(0.699\), \(0.721\), and \(0.724\), respectively.  
These findings suggest that even minimal manual supervision significantly boosts performance, validating DocAnnot’s effectiveness in reducing human annotation effort while maintaining high-quality model outcomes.

\begin{table}[htbp]
\centering
\caption{Hybrid Annotation Performance on the CORD Dataset: UDOP trained with auto-annotated data (DocAnnot) combined with increasing fractions of human-annotated samples. Results show consistent improvements in precision, recall, and F1 score with limited manual supervision.}
\label{tab:hybrid_annotation}
\resizebox{\textwidth}{!}{
\begin{tabular}{|l|l|c|c|c|}
\hline
Model & Training Data & Precision & Recall & F1 Score \\
\hline
\multirow{4}{*}{UDOP} & Auto-Annotated (DocAnnot only) & 0.7364 & 0.6424 & 0.6861 \\
                      & + 10\% Human-Annotated         & 0.7521 & 0.6530 & 0.6991 \\
                      & + 20\% Human-Annotated         & 0.7712 & 0.6784 & 0.7218 \\
                      & + 30\% Human-Annotated         & 0.7560 & 0.6948 & 0.7241 \\
\hline
\end{tabular}
}
\end{table}

\section{Conclusion}
Our paper presents DocAnnot, a Multimodal GenAI-Powered Automated Annotation Framework designed to accelerate the creation of training datasets for KIE models. By leveraging an LVLM, OCR, and the novel SICM algorithm, DocAnnot streamlines and automates the document annotation process. Our experimental study demonstrates the effectiveness of DocAnnot in producing high-quality datasets for training KIE models without human intervention. The ablation study further highlights the importance of SICM in ensuring reliable label-value associations. While models trained on fully human-annotated data still achieve the highest performance, DocAnnot offers a practical and scalable alternative that significantly reduces annotation time and cost. Furthermore, our hybrid annotation experiments show that incorporating even a small fraction of human-labeled data (10–30\%) alongside DocAnnot outputs can substantially boost model precision, achieving F1 scores comparable to human-supervised training. This positions DocAnnot as a valuable component in real-world annotation pipelines, especially in resource-constrained or rapid development scenarios.

Future work will focus on expanding the evaluation to more complex and diverse real-world datasets to better assess generalizability. We also plan to evaluate DocAnnot on synthetic out-of-distribution documents to better understand the effects of LVLM pretraining overlap and to further test the adaptability of the framework.

\bibliographystyle{splncs04}

\bibliography{references}





\end{document}